\documentclass[runningheads]{llncs}
\usepackage{amsmath,amssymb,amsfonts}
\usepackage{subcaption}
\usepackage{graphicx}
\usepackage{hyperref}
\usepackage{arydshln}
\usepackage{multirow}
\usepackage[export]{adjustbox}

 \providecommand{\norm}[1]{\lVert#1\rVert}

\begin{document}
\title{Patch-wise Deep Metric Learning for Unsupervised Low-Dose CT Denoising
\thanks{This study was approved by the institutional review board(IRB) of our institution, and the requirement for informed consent was waived. 
	This study was supported by the National Research Foundation of Korea(NRF) grant NRF-2020R1A2B5B03001980, KAIST Key Research Institute (Interdisciplinary Research Group) Project, Field-oriented Technology Development Project for Customs Administration through NRF funded by the Ministry of Science \& ICT(MSIT) and Korea Customs Service(NRF-2021M3I1A1097938). Sunkyoung You was supported by the Bio \& Medical Technology Development Program of the NRF \& funded by the MSIT (No.NRF-2019M3E5D1A02068564)}
}

%
%

\author{Chanyong Jung\inst{1} \and
 Joonhyung Lee\inst{1}\thanks{Joonhyung Lee is currently at VUNO Corp.} \and
 Sunkyoung You\inst{2} \and
 Jong Chul Ye\inst{1,3}}

\authorrunning{C. Jung et al.}


 \institute{ Department of Bio and Brain Engineering, Korea Advanced Institute of Science and Technology (KAIST), Daejeon, Republic of Korea \and
 Department of Radiology, Chungnam National University College of Medicine and Chungnam National University Hospital, Daejeon, Republic of Korea \and
 Kim Jaechul Graduate School of AI, KAIST, Daejeon, Republic of Korea }

\maketitle              

\begin{abstract}
The acquisition conditions for low-dose and high-dose CT images are usually different, so that the shifts in the CT numbers often occur. Accordingly, unsupervised deep learning-based approaches, which learn the target image distribution, often introduce CT number distortions and result in detrimental effects in diagnostic performance. To address this, here we propose a novel unsupervised learning approach for lowdose CT reconstruction using patch-wise deep metric learning. The key idea is to learn embedding space by pulling the positive pairs of image patches which shares the same anatomical structure, and pushing the negative pairs which have same noise level each other. Thereby, the network is trained to suppress the noise level, while retaining the original global CT number distributions even after the image translation. Experimental results confirm that our deep metric learning plays a critical role in producing high quality denoised images without CT number shift.

\end{abstract}
\section{Introduction}
\label{sec:introduction}
In computed tomography (CT), multiple X-ray projection images are obtained at various angles, which incurs considerable radiation exposure to a patient\cite{Brenner2007}. In addition, for a temporal CT imaging to evaluate the pathologies of the ears especially for children, it is practical standard to use low-dose CT. Thus, low-dose CT reconstruction, which reduces the radiation dose for each projection while maintaining the image quality, is an important research topic.

Recently, deep learning approaches have become a main approach for low-dose CT reconstruction  \cite{Tang2019LDCT,Chen2017LDCT}, which has resulted many commercially available products\cite{truefidelity}. Furthermore, the difficulty of obtaining matched CT data pairs has led to exploring various unsupervised learning approaches \cite{Yuan2020LDCT,Bai2021LDCT}.
In particular, image translation methodology is successfully employed for the low-dose CT noise suppression by learning the noise patterns through a comparison between the low-dose CT(LDCT) and high-dose CT(HDCT) distributions \cite{LDCT_WGAN,gu2021cyclegan,ot-cyclegan}. 

Despite the success, the approach is not free of limitation. In particular, for the real-world dataset, there often exists CT number shift between LDCT and HDCT dataset due to the different acquisition conditions. Hence, the networks often result the distortion in pixel values. 
Unfortunately, the change of the CT numbers has detrimental effects to the radiological diagnostic performance, since it is used as a diagnostic indicator \cite{CTshift}.
Especially in the temporal bone CT scans which we are interested in this paper, 
the CT number supports the evaluation of a pathologies of the ear, such as an inflammation within the inner ear, and a cholesteatoma in early stage \cite{tempCT2015,HU2011,HU2014}.
Wavelet-assisted approaches \cite{gu2021adain,gu2021cyclegan} partially address this problem by constraining the pixel value variations only in high-frequency subbands.
However, even in the high-frequency image, there exists many important structural information, which can be altered during the image translation so that the localized CT number variation is unavoidable. 

Recently, image translation methods utilizing patch-wise interaction have received significant attention\cite{cut,negcut}. By the contrastive loss, the mutual information(MI) between the patches are maximized, pulling positive pairs and pushing negative pairs. However, for denoising tasks, maximizing MI is not appropriate, since a preservation of noise pattern also contributes to the increase of MI. 

Here, we propose a novel approach for the LDCT denoising, suggesting the pull-push mechanism for the denoising process by newly defined pairs and a metric. We set the positive pair as the patches from same location of noisy input and denoised output, which share anatomic structure. Also, the negative pair is set by the patches from different location of same image, which have same noise level. Thereby, the network is trained to maintain the structural information, and suppress the noise level. The experimental results on the synthetic noise dataset and a real-world dataset verify the effectiveness of the proposed method.

\section{Method}\label{sec:method}
\subsection{Overview of the proposed method}
\begin{figure}[t]
	\centering 
	\includegraphics[width=0.95\textwidth]{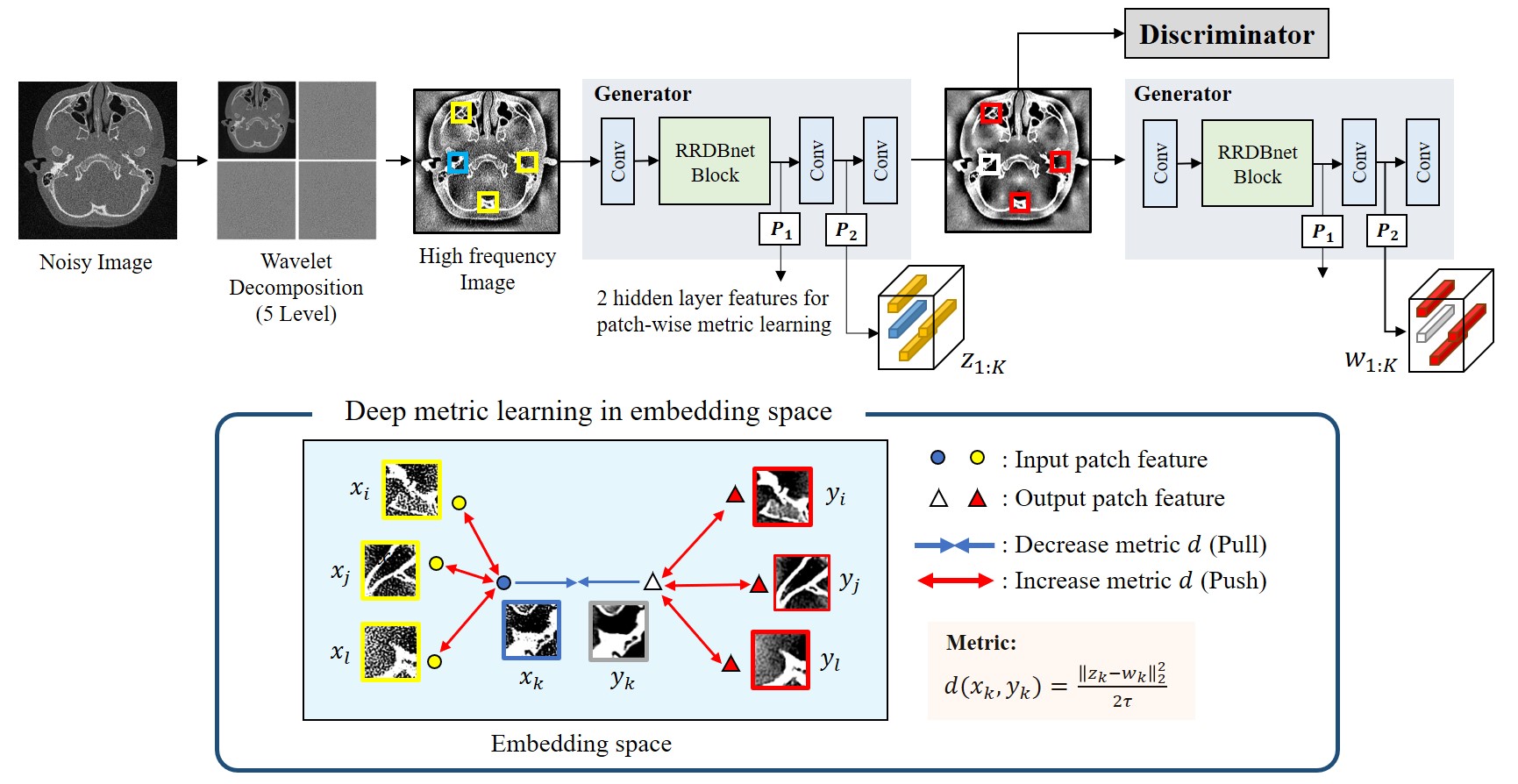}
	\caption{Proposed deep metric learning for low-dose CT denoising. The positive pair $x_k$, $y_k$ pull each other to decrease the metric $d$ between them. $x_i, x_j, x_l$ are pushed away $(i, j, l \neq k)$ from $x_k$, as they have same noise level and different contents. Likewise, $y_i, y_j, y_l$ are pushed away from $y_k$.}
	\label{fig:proposed_concept}

\end{figure}

In the proposed framework, we aim to learn the embedding space which focus on the anatomic structure features, and discards the noise features. In the training process, the image patches with same contents and different noise level will pull each other, to learn the shared feature between them (i.e. anatomic structure). On the other hand, the image patches with same noise level and different contents will push each other, to suppress the network to learn the shared feature (i.e. noise). Deep metric is utilized for the pull and push between the image patches.

Specifically, as shown in Fig. \ref{fig:proposed_concept}, we first obtain a high-frequency image by a wavelet transform, and 
then exploit the patch-wise deep metric learning using the features of the image before and after the denoising process.
Here, similar to \cite{gu2021adain,gu2021cyclegan},
high frequency images are obtained by a wavelet decomposition, putting zeros at the low-frequency band, and then performing wavelet recomposition. This preprocessing preserves the low frequency image of input.

Then, the high-frequency images from the input and the output images
are processed by shared feature extractors with projection heads to generate
patch-by-patch latent vectors. In addition,
as for the deep metric learning of these latent vectors, we propose a loss so that effective representation for the image generation can be obtained. The final denoised image is obtained by adding the generator output with low freqeuncy image of input.  

Here, we discuss the relation of our work with the contrastive learning. The goal of the contrastive loss as suggested in the related work\cite{cut} is to maximize the MI between the patches of input and output images. Hence, the negative pair is defined as the patches from input and output with different spatial location, to implement the sampling from marginal distributions. Then, the loss pushes the pairs with different noise level, which leads to different solution with our loss function. Both loss functions utilize the pulling and the pushing of the image patches, however, the objective and the pull-push mechanism of each loss are different. Therefore, we propose our work as deep metric learning which generalize the methods based on the pull-push mechanism.

\subsection{Patch-wise deep metric learning}
As shown in Fig.\ref{fig:proposed_concept}, we obtain the embedding space by the pull and push between the image patches. Let $x_k\in \mathcal{X}$ and $y_j\in \mathcal{Y}$ denote the image patches from the input and denoised output image, respectively, where the $k,j$ are the indices for the patch locations, and $\mathcal{X}$ and $\mathcal{Y}$ are domains for LDCT and HDCT images. Also, let $z_k$ and $w_k$ are the L2 normalized embedding vectors for $x_k$ and $y_k$.
Then, we define the metric $d$ as following:
\begin{equation}
    d(x_k, y_k) = \frac{\left\| z_k - w_k \right\|_2^2 }{2\tau } 
\end{equation}
where $\tau$ is the temperature parameter. 

Using the defined metric $d$, we decrease the distance between the positive pairs, and make the negative pairs be far apart in the embedding space. We optimize the following metric loss $\mathcal{L}_m$ given as:
\begin{align}
     \mathcal{L}_m := \mathbb{E}_p\left[\exp( d(x_k, y_k))\right] - \mathbb{E}_{q_n}\left[\exp (d(x_k, x_j))\right] - \mathbb{E}_{q_o}\left[\exp(d(y_k, y_j))\right]
\end{align}
where $p$ is the distribution for the positive pair, $q_n$ is for the negative pair from the input image, and $q_o$ is for the negative pair from the output image. 
Since $||z_k-w_k||_2^2 = 2-2z_k^\top w_k$, the loss function can be rewritten as:
\begin{equation}
  \mathcal{L}_m := - \mathbb{E}_p[\exp (z_k^\top w_k / \tau)] + \mathbb{E}_{q_n}[\exp (z_k^\top z_j/ \tau)] + \mathbb{E}_{q_o}[\exp (w_k^\top w_j/ \tau)]
\end{equation}

\subsection{Network architecture}

\begin{figure}[t]
	\centering 
	\includegraphics[width=1\textwidth]{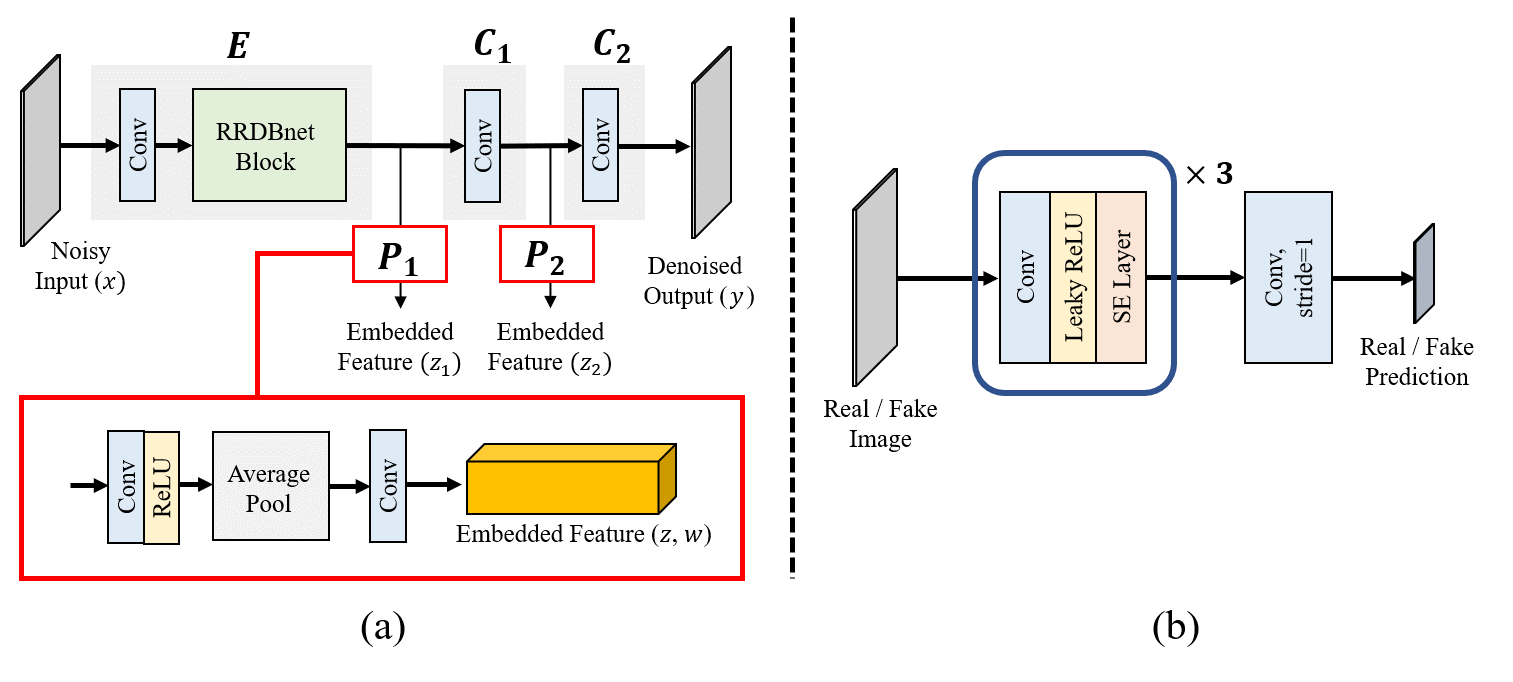}
	
	\caption{Structure of the generator $G$ and discriminator $D$. Two intermediate features of the $G$ are used for deep metric learning. (a) $G$ consists of feature extractor $E$, and convolutional layers $C_1, C_2$ and projection heads $P_1, P_2$. (b) $D$ is composed of three repeated convolutional blocks.}
	\label{fig:net}
\end{figure}

Our generator architecture is shown in Fig.~\ref{fig:net}, where
the generator $G$ is composed of several blocks which are encoder network $E$ and convolutional layers $C_1, C_2$:
\begin{align}\label{eq:G}
G: x\in \mathcal{X} \mapsto \mathcal{Y},&\quad\mbox{where}\quad G(x) = (C_2 \circ C_1 \circ E)(x)
\end{align}
where $\circ$ denotes the composite function. The encoder
network $E$ is composed of convolutional layer followed by RRDBnet \cite{rrdb} architecture.
Furthermore, we use two latent feature extractors:
\begin{align*}
F_1(x) &= (P_1 \circ E)(x)\\
F_2(x) &= (P_2 \circ C_1 \circ E)(x)
\end{align*}
where $P_1, P_2$ are projection heads shown in Fig.~\ref{fig:net}.

Then, the overall loss function is given as:
\begin{align}
    \mathcal{L}=\mathcal{L}_{GAN} + \lambda_{idt} \mathcal{L}_{idt} +  \lambda_{m} \mathcal{L}_{m}
\end{align}
where $\mathcal{L}_{GAN}, \mathcal{L}_{idt}$ and $\mathcal{L}_{m}$ are the GAN loss, identity loss and deep metric loss, respectively, 
and $\lambda_{idt}$ and $\lambda_{m}$ and denote their associated weighting parameters. 
 As for the GAN loss $\mathcal{L}_{GAN}$, we use the LSGAN loss\cite{Mao2017}:
\begin{align}
\mathcal{L}_{GAN}(E,\{C_i\}, D) = \mathcal{L}_{GAN}(E,\{C_i\}) + \mathcal{L}_{GAN}(D)
\end{align}
where
\begin{align}
\mathcal{L}_{GAN}(G) =~& \mathbb{E}_{x\sim P_X}\left[\norm{D\left(G(x)\right)-1}_2\right]\\
\mathcal{L}_{GAN}(D) =~& \mathbb{E}_{y\sim P_Y}\left[\norm{D\left(y\right)-1}_2\right] +\mathbb{E}_{x\sim P_X}\left[{\norm{D\left(G(x)\right)}}_2\right]
\end{align}
The identity loss is used to impose the reconstruction constraint for the target domain image on the network, thereby preventing any artifact of the output image caused by GAN loss. The idt loss $\mathcal{L}_{idt}$ is given as:
 \begin{equation}
\mathcal{L}_{idt}\left(E,\{C_i\}\right) 
=\mathbb{E}_{y\sim P_Y}\left[\norm{(C_2 \circ C_1 \circ E)(y)-y}_1\right],
\label{eq:identity_loss}
\end{equation}
Finally, the deep metric loss $\mathcal{L}_m$ is composed of two terms, since we use two intermediate features as shown in Fig.~\ref{fig:net}
\begin{align}
\mathcal{L}_{m}\left(E,C_1,\{P_i\}\right)=\mathcal{L}_{m,1}\left(E,P_1\right) +\mathcal{L}_{m,2}\left(E,C_1,P_2\right) 
\end{align}
 
\section{Experiments}\label{sec:exp}


We verify our method using two datasets. First, we use the publicly released CT dataset from {the 2016 NIH-AAPM-Mayo Clinic Low Dose CT Grand Challenge} dataset (AAPM), and the real-world temporal CT scan dataset. The real-world temporal CT scans are obtained at 100kVp. Collimation is $128mm \times 0.6mm$, the pitch is set to 1, and the slice thickness is 0.6mm. 
Reference mAs for low-dose and high-dose scans are 46mAs and 366mAs, respectively. H70s kernel for low-dose and H60s kernel high-dose scans are used. The dataset consists of 54 volumes of low-dose, and  34 volumes of high-dose scans. We split the low-dose data into two sets, a train set with 45 volumes and a test set with 9 volumes. 

We compare our method with existing denoising methods based on the image translation framework, such as cycleGAN\cite{Zhu2017}, wavCycleGAN\cite{gu2021cyclegan}, GAN-Circle\cite{gan-circle}and Cycle-free invertible cycleGAN\cite{invertible-cyclegan}. We measure PSNR, SSIM to compare the denoising performance for AAPM dataset. For temporal CT scans, the ground truth is not accessible. Hence, we measure the mean and standard deviation of pixels to investigate the CT number shift and denoising performance.  

The method is implemented by the PyTorch \cite{NIPS2019_9015} with two NVIDIA GeForce GTX 2080Ti GPU devices.
The Adam optimizer \cite{Kingma2014} is used with $\beta_1=0.5$ and $\beta_2=0.999$. The learning rate is scheduled to be constant until the half of the total training epoch, and then linearly decreased to zero.
The inputs for the network are randomly cropped images with size of 128$\times$128. The projection heads downsample the features by the 2$\times$2 pooling layer for temporal CT dataset. In case of AAPM dataset, the projection heads do not downsample the feature. The source code is available at: \texttt{https://github.com/jcy132/DML\_CT}

\subsection{AAPM dataset}

\begin{figure*}[t]

	\centering 
	\includegraphics[width=1\linewidth]{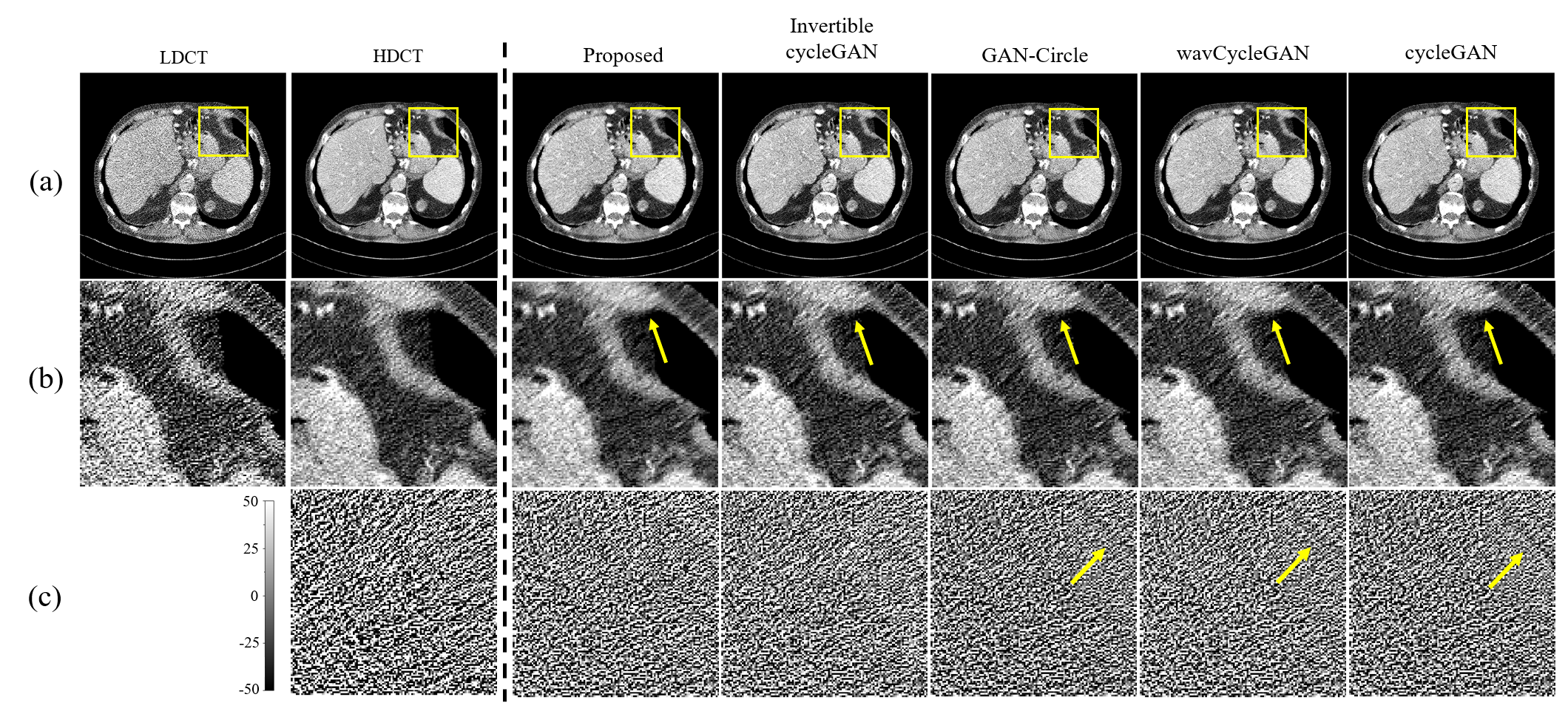}
	\caption{Results from AAPM dataset. (a) CT images for the comparison, displayed with (-150HU, 210HU). (b) Zoomed image for the comparison of fine structures. (c) Difference images of zoomed area subtracted by the low-dose images, displayed with (-50HU, 50HU).}
	\label{fig:result_AAPM_img}

\end{figure*}

\begin{table}[t]
	\caption{Quantitative metrics (PSNR, SSIM) to compare the performance with the previous methods on AAPM dataset.}
	\label{table:result_AAPM}
	\begin{center}
		\resizebox{\textwidth}{!}{
			{	
				\small
				\renewcommand{\arraystretch}{1.2}
				\begin{tabular}{@{\extracolsep{5pt}}p{1.3cm}|p{1.2cm}:ccccc}
					\hline
					\textbf{} & LDCT & Proposed & Invertible-cycleGAN  & GAN-Circle & wavCycleGAN & cycleGAN \\ \hline 
					\textbf{PSNR} & 33.300 & \textbf{38.110} & 37.855 & 37.795 & 37.736 & 37.673 \\ 
					\textbf{SSIM} & 0.740  & \textbf{0.875} & 0.867  & 0.867 & 0.867  & 0.865 \\  \hline
				\end{tabular}
			}
		}
	\end{center}
\end{table}

For AAPM dataset, the network was  trained with learning rate 2e-4 with epoch 200. The wavelet decomposition is proceeded by level 6 using db3 wavelet filter. The batch size is 8, and the $\tau=0.15$ for the metric $d$. $\lambda_{idt}=5$ and $\lambda_{m}=0.1$. We used 3112 images for training, and evaluated for 421 test images. 

Fig.~\ref{fig:result_AAPM_img} shows the visual quality of the output images. Compared to the previous works, the proposed method suppressed the noise level in the output image, while preserving the structures. The difference images indicates that the HU value is also well preserved during the denoising process of our method. Moreover, our method shows less distortion in pixel values, compared to the other methods which reveals the structures in the difference images. 
The quantitative result also supports the superiority of the proposed method. As shown in Table \ref{table:result_AAPM}, the proposed method shows the best performance in PSNR and SSIM, which indicates the successful denoising performance compared to other methods.

\subsection{Temporal CT scans}
   
For Temporal CT dataset, the networks are trained with learning rate 2e-4 with epoch 100. Wavelet decomposition is proceeded by level 5 with db3 wavelet filter. The batch size is 4, and the $\tau=0.12$ for the metric $d$. $\lambda_{idt}=5$ and $\lambda_{m}=0.1$.

\begin{figure*}[t]
	\centering 
	\includegraphics[width=0.95\linewidth]{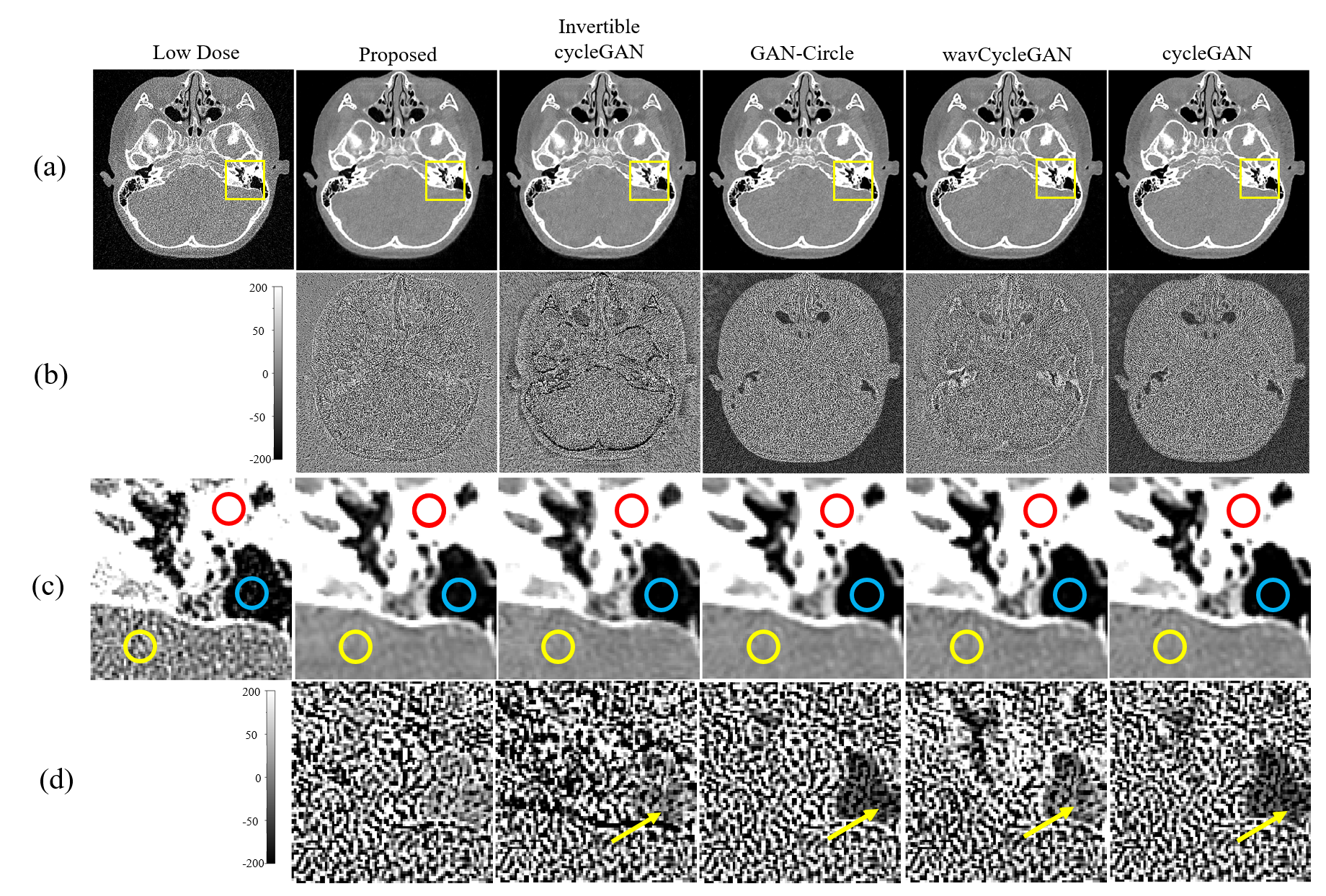}
	\caption{Results from the temporal CT dataset. (a) CT images for the comparison, displayed with (-780HU, 820HU). (b) Difference images, displayed with (-200HU, 200HU). (c) Zoomed image for the temporal region. Colored circles are the selected regions for the quantitative evaluation. (d) Difference image for the zoomed images. Yellow arrows indicate the distortion of the CT numbers.}
	\label{fig:result_tempCT_img}

\end{figure*}

\begin{table}[t]
	\caption{Quantitative comparison for the temporal CT data. The mean values of air and brain area shows the CT number shift problem between the LDCT and HDCT images. Our goal is to produce the outputs with similar mean with LDCT but low Std value.}
	
	\label{table:img_patch_HU}
	\begin{center}
		\resizebox{\textwidth}{!}{
			{	
				\small
				\renewcommand{\arraystretch}{1.2}
				\begin{tabular}{@{\extracolsep{10pt}}ccccccccccc}
					\hline
					& \multicolumn{4}{c}{\textbf{Image patches with Homogeneous media}} & \multicolumn{6}{c}{\textbf{Selected temopral region}} \\ 
					\cline{2-5}
					\cline{6-11}
					&\multicolumn{2}{c}{Brain}&\multicolumn{2}{c}{Air}&\multicolumn{2}{c}{Soft Tissue }&\multicolumn{2}{c}{Cavity}&\multicolumn{2}{c}{Bone} \\ 
					\cline{2-3}
					\cline{4-5}
					\cline{6-7}
					\cline{8-9}
					\cline{10-11}
					& Mean & Std & Mean &Std &Mean & Std & Mean & Std & Mean & Std \\
					\hline
					LDCT & \textbf{35.02} & 289.47 & \textbf{-845.28} & 132.82 & \textbf{30.37} & 285.97 & \textbf{-775.71} & 182.86 & \textbf{1350.72} & 344.54 \\ 
					HDCT & 37.36 & \textbf{48.62} & -972.29 & \textbf{32.95}  & N/A & N/A & N/A & N/A& N/A & N/A \\  \hdashline
					cycleGAN & 36.78 & 59.03 & -968.14 & 40.12 & 38.06 & 52.89 & -923.45 & 36.47 & 1389.07 & 116.46 \\ 
					WavCycleGAN & 32.69 & 60.98 & -852.94 & 33.13 & 49.71 & 58.43 & -859.92 & 38.26 & 1375.65 & 156.43 \\
					GAN-Circle & 39.43 & 49.23 & -994.40 & 38.11 & 40.29 & 50.22 & -925.42 & 36.50 & 1364.28 & 115.47 \\
					Invertible-cycleGAN & 37.74 & 53.46 & -846.87 & 33.47 & 44.44 & 54.44 & -801.20 & 44.12 & 1384.13 & 111.81 \\ \hdashline
					Proposed & \textbf{35.05} & \textbf{49.06} & \textbf{-845.65} & \textbf{32.99} & \textbf{29.39} & \textbf{48.15} & \textbf{-777.63} & \textbf{33.65} & \textbf{1352.57} & \textbf{72.98} \\ \hline
				\end{tabular}
			}
		}
	\end{center}

\end{table}

From the output and difference images in Fig.~\ref{fig:result_tempCT_img}, we can see that 
the proposed method shows improved denoising output which has less distortion in the HU values. Also, the proposed method preserved the bone structure more than other methods.
In contrast to our method, cycleGAN and GAN-Circle output the images with distorted HU values as shown in difference image. Since the methods translates the image to be similar to the target domain, it is vulnerable to the HU value shift problem caused by the different acquisition condition between the low-dose and high-dose scans. 
Despite the wavelet-domain processing, invertible-cycleGAN and wavCycleGAN produce the artifacts at the inner ear shown in the difference images in Fig.~\ref{fig:result_tempCT_img}, which are undesirable for the temporal CT data with fine structures in the temporal region.

For the quantitative evaluation of the methods, we compare the mean and standard deviation (Std) values of the pixels, to investigate the HU value shift and denoising performance. 
We proceed the quantitative evaluation in two different ways. First, we crop 60$\times$60 sized 300 image patches with homogeneous media (i.e. air and brain), and obtain the average value of the mean and std. The CT number shift is investigated by the mean, and the denoising performance is compared by the std.
Next, we compare the methods for the temporal region. We select the soft tissue, cavity and bone area, shown as yellow, blue and red circles in Fig.~\ref{fig:result_tempCT_img}(c), respectively. Then, we measure the mean and std.

We present the results in Table \ref{table:img_patch_HU}.
The proposed method shows the lowest Std values which indicate the successful denoising performance. Also, the mean values are similar to the LDCT, which verifies a robustness to the CT number shift problem of the real-world dataset.
The cycleGAN and GAN-Circle show a vulnerability to the CT number shift problem. Specifically, the methods output the similar mean values with the HDCT, distorting the pixel values of the input image. The distortion is also observed for the mean values of the temporal region, which is coherent with Fig.~\ref{fig:result_tempCT_img}. 
The wavCycleGAN and invertible-cycleGAN are somewhat robust to the CT number shift problem, since the methods utilize the high-frequency images. However, the CT number shift  at the inner ear degrades the performance. The results for the temporal region in Table \ref{table:img_patch_HU} reveal the distortion of the CT number and the degradation of the denoising performance, which is coherent with the difference images in Fig.~\ref{fig:result_tempCT_img}.

\section{Conclusion}\label{sec:conclusion}
In this paper, we developed a novel denoising methodology based on the patch-wise deep metric learning. In our framework, the patches sharing the same anatomic structure is used as positive pair, and patches which have similar noise level but different spatial information are considered as negative samples. By the push and the pull between the features in the embedding space, the network focuses on the anatomic information and neglect the noise features. 
The results verified that the proposed method is effective technique to improve the denoising performance without the CT number distortion.

%
%
\bibliographystyle{splncs04}
\bibliography{paper1843.bib}

\begin{thebibliography}{10}
\providecommand{\url}[1]{\texttt{#1}}
\providecommand{\urlprefix}{URL }
\providecommand{\doi}[1]{https://doi.org/#1}

\bibitem{HU2014}
Ahn, S.H., Kim, Y.W., Baik, S.K., Hwang, Yeon, J., Lee, I.W.: Usefulness of
  computed tomography hounsfield unit measurement for diagnosis of congenital
  cholesteatoma. Journal of the Korean Radiological Society  \textbf{70},
  153--158 (2014)

\bibitem{Bai2021LDCT}
Bai, T., Wang, B., Nguyen, D., Jiang, S.: Probabilistic self-learning framework
  for low-dose {CT} denoising. Medical Physics  \textbf{48}(5),  2258--2270
  (2021)

\bibitem{Brenner2007}
Brenner, D.J., Hall, E.J.: {Computed tomography - An increasing source of
  radiation exposure} (nov 2007). \doi{10.1056/NEJMra072149},
  \url{http://www.nejm.org/doi/abs/10.1056/NEJMra072149}

\bibitem{Chen2017LDCT}
Chen, H., Zhang, Y., Kalra, M.K., Lin, F., Chen, Y., Liao, P., Zhou, J., Wang,
  G.: {Low-dose CT With a residual encoder-decoder convolutional neural
  network}. IEEE Transactions on Medical Imaging  \textbf{36}(12),  2524--2535
  (2017)

\bibitem{tempCT2015}
CL, T., A, S., S, S., AS, S., K., S.: Role of high resolution computed
  tomography in evaluation of pathologies of temporal bone. Journal of Clinical
  and Diagnostic Research  \textbf{9},  TC07–TC10 (2015)

\bibitem{CTshift}
Cruz-Bastida, J.P., Zhang, R., Gomez-Cardona, D., Hayes, J., Li, K., Chen,
  G.H.: Impact of noise reduction schemes on quantitative accuracy of ct
  numbers. Medical Physics  \textbf{46}(7),  3013--3024 (2019)

\bibitem{gu2021cyclegan}
Gu, J., Yang, T.S., Ye, J.C., Yang, D.H.: {CycleGAN denoising of extreme
  low-dose cardiac CT using wavelet-assisted noise disentanglement}. Medical
  Image Analysis  \textbf{74},  102209 (2021)

\bibitem{gu2021adain}
Gu, J., Ye, J.C.: {AdaIN-based tunable cycleGAN for efficient unsupervised
  low-dose CT denoising}. IEEE Transactions on Computational Imaging
  \textbf{7},  73--85 (2021)

\bibitem{truefidelity}
Healthcare, G.: {TrueFidelity CT. Technical white paper on deep learning image
  reconstruction},
  \url{https://www.gehealthcare.co.kr/products/computed-tomography/truefidelity}

\bibitem{Kingma2014}
Kingma, D.P., Ba, J.: {Adam: A Method for Stochastic Optimization}. In: Bengio,
  Y., LeCun, Y. (eds.) 3rd International Conference on Learning
  Representations, (ICLR) 2015, May 7-9, 2015, Conference Track Proceedings.
  San Diego, CA, USA (2015), \url{http://arxiv.org/abs/1412.6980}

\bibitem{invertible-cyclegan}
Kwon, T., Ye, J.C.: Cycle-free cyclegan using invertible generator for
  unsupervised low-dose ct denoising. IEEE Transactions on Computational
  Imaging  \textbf{7},  1354--1368 (2021)

\bibitem{Mao2017}
Mao, X., Li, Q., Xie, H., Lau, R.Y., Wang, Z., Smolley, S.P.: {Least Squares
  Generative Adversarial Networks}. In: Proceedings of the IEEE International
  Conference on Computer Vision. vol. 2017-October, pp. 2813--2821. Institute
  of Electrical and Electronics Engineers Inc. (dec 2017).
  \doi{10.1109/ICCV.2017.304},
  \url{https://ieeexplore.ieee.org/document/8237566}

\bibitem{HU2011}
Park, M.H., Rah, Y.C., Kim, Y.H., hoon Kim, J.: Usefulness of computed
  tomography hounsfield unit density in preoperative detection of cholesteatoma
  in mastoid ad antrum. American Journal of Otolaryngology  \textbf{32}(3),
  194--197 (2011)

\bibitem{cut}
Park, T., Efros, A.A., Zhang, R., Zhu, J.Y.: Contrastive learning for unpaired
  image-to-image translation. In: ECCV (2020)

\bibitem{NIPS2019_9015}
Paszke, A., Gross, S., Massa, F., Lerer, A., Bradbury, J., Chanan, G., Killeen,
  T., Lin, Z., Gimelshein, N., Antiga, L., Desmaison, A., Kopf, A., Yang, E.,
  DeVito, Z., Raison, M., Tejani, A., Chilamkurthy, S., Steiner, B., Fang, L.,
  Bai, J., Chintala, S.: {PyTorch: An Imperative Style, High-Performance Deep
  Learning Library}. In: Wallach, H., Larochelle, H., Beygelzimer, A., d'Alché
  Buc, F., Fox, E., Garnett, R. (eds.) Advances in Neural Information
  Processing Systems 32, pp. 8026--8037. Curran Associates, Inc. (2019)

\bibitem{ot-cyclegan}
Sim, B., Oh, G., Kim, J., Jung, C., Ye, J.C.: Optimal transport driven cyclegan
  for unsupervised learning in inverse problems. SIAM Journal on Imaging
  Sciences  \textbf{13}(4),  2281--2306 (2020)

\bibitem{Tang2019LDCT}
Tang, C., Li, J., Wang, L., Li, Z., Jiang, L., Cai, A., Zhang, W., Liang, N.,
  Li, L., Yan, B.: Unpaired low-dose {CT} denoising network based on
  cycle-consistent generative adversarial network with prior image information.
  Computational and Mathematical Methods in Medicine  \textbf{2019} (2019)

\bibitem{negcut}
Wang, W., Zhou, W., Bao, J., Chen, D., Li, H.: Instance-wise hard negative
  example generation for contrastive learning in unpaired image-to-image
  translation. In: Proceedings of the IEEE/CVF International Conference on
  Computer Vision (ICCV). pp. 14020--14029 (October 2021)

\bibitem{rrdb}
Wang, X., Yu, K., Wu, S., Gu, J., Liu, Y., Dong, C., Qiao, Y., Loy, C.C.:
  {ESRGAN: Enhanced super-resolution generative adversarial networks}. In: The
  European Conference on Computer Vision Workshops (ECCVW) (2018)

\bibitem{LDCT_WGAN}
Yang, Q., Yan, P., Zhang, Y., Yu, H., Shi, Y., Mou, X., Kalra, M.K., Zhang, Y.,
  Sun, L., Wang, G.: {Low-Dose CT image denoising using a generative
  adversarial network with wasserstein distance and perceptual loss}. IEEE
  Transactions on Medical Imaging  \textbf{37}(6),  1348--1357 (2018).
  \doi{10.1109/TMI.2018.2827462}

\bibitem{gan-circle}
You, C., Li, G., Zhang, Y., Zhang, X., Shan, H., Li, M., Ju, S., Zhao, Z.,
  Zhang, Z., Cong, W., Vannier, M.W., Saha, P.K., Hoffman, E.A., Wang, G.: Ct
  super-resolution gan constrained by the identical, residual, and cycle
  learning ensemble (gan-circle). IEEE Transactions on Medical Imaging
  \textbf{39}(1),  188--203 (2020)

\bibitem{Yuan2020LDCT}
Yuan, N., Zhou, J., Qi, J.: Half2half: deep neural network based {CT} image
  denoising without independent reference data  \textbf{65}(21),  215020 (nov
  2020). \doi{10.1088/1361-6560/aba939},
  \url{https://doi.org/10.1088/1361-6560/aba939}

\bibitem{Zhu2017}
Zhu, J.Y., Park, T., Isola, P., Efros, A.A.: {Unpaired Image-to-Image
  Translation Using Cycle-Consistent Adversarial Networks}. In: Proceedings of
  the IEEE International Conference on Computer Vision. vol. 2017-Octob, pp.
  2242--2251. Institute of Electrical and Electronics Engineers Inc. (dec
  2017). \doi{10.1109/ICCV.2017.244}

\end{thebibliography}

\end{document}